# A reconstruction method for binary limited-data tomography using a dictionary-based sparse shape recovery

Haytham A. Ali

Hiroyuki Kudo
University of Tsukuba

# A reconstruction method for binary limited-data tomography using a dictionary-based sparse shape recovery

Haytham A. Ali[*1], Katsuya Fujii[*] and Hiroyuki Kudo[*]

[*] *Information and Systems, University of Tsukuba, Tsukuba, Ibaraki 305-8573, Japan*

[1]*Department of Mathematics, Faculty of Science, Sohag University, Sohag 82524, Egypt*

**Abstract**： Binary tomography is concerned with reconstructing a binary image from a very small number or other limited CT projection data. This problem itself not only possesses several medical imaging applications but also can be considered a model of general inverse problems to recover the object shape from limited measured data. Several approaches such as the Mumford-Shah method and various level-set methods have been investigated, but most of them lead to a non-convex optimization due to the difficulty to handle the binary constraint. We propose a new method based on a convex optimization inspired by dictionary-based shape recovery. In the proposed method, the object boundary of the binary image is represented by a level set of linear combinations of basis vectors in the dictionary. Using the dictionary, the object boundary is reconstructed by finding weights of the linear combination that best match the measured data. We create the dictionary by using the Gaussian radial basis function (GRBF). More concretely, we use Gaussian functions as a basis function placed at sparse grid points to represent the parametric level-set function and provide more flexibility in the binary representation of the reconstructed image. The simulation results of CT image reconstruction from only four projection data demonstrate that the proposed method can recover the object boundary more accurately compared with other competitive methods. The significance of our approach is the formulation with a tractable convex program while keeping moderate mathematical rigorousness.

## 1. Introduction

Tomography imaging is known as a method for displaying precise details inside the scanned object, i.e., it visualizes the internal structures of objects and thus has a wide range of applications such as medicine, science, industry, and electron tomography [1-3]. In all these applications that deal with reconstructing images from a given set of projection data, it is highly desirable to decrease the number of rays penetrating through the unknown object. Mathematically, this object can be represented as a function with a domain that can be discrete or continuous. Discrete Tomography (DT) deals with reconstructing discrete images consisting of a few different materials by using a limited number of projections. Recently, Batenburg and Sijbers have developed an algorithm called Discrete Algebraic Reconstruction Technique (DART) for DT, which provides high-quality reconstructions [3]. Even though this technique has its advantages, it requires more computation time, which limits its use in practical applications.

## 2. Problem Formulation

For binary image reconstruction, we can use the following linear system of equations

$$b = Au, \qquad (1)$$

where $A \in \mathbb{R}^{M\times N}$ is the measurement matrix, $b \in \mathbb{R}^{M}$ is a measurement and the image $u \in \mathbb{R}^{N}$.

To solve (1), we used a regularized least-squares problem

$$\min\ f(u) = \|Au - b\|_2^2 + \mathcal{L}(u) \quad , \qquad (2)$$

where $\mathcal{L}(u)$ is the regularization term. Based on the properties of the problem we can choose $\mathcal{L}$. This regularization term is used to enforce smoothness and boundedness of $u$. In our proposed the number of parameters involved in the problem is very small. As a result, the underdetermined problem can be made overdetermined, and the problem becomes better posed. According to this the parametrization idea is empirically found to be well-posed enough that no necessary regularization terms need to be added to the cost function. So, Eq. (2) can be written as

$$\min\ f(u) = \|Au - b\|_2^2 \quad , \qquad (3)$$

This optimization problem (3) will solve by using a Newton like algorithm [4] as

$$u_{k+1} = u_k - \lambda_k \mathcal{H}_k^{-1} \nabla f(u_k),$$

where $\lambda_k$ denotes the step-size parameter, $\mathcal{H}_k$ denotes an approximation of the Hessian matrix of $f$ calculated at $u = u_k$, and $\nabla f(u_k)$ is the gradient of $f$ at $u_k$.

## 3. Level-set methods

By using the parametric level-set function $f(x,\alpha)$, we can express the image $u$ as [5]

$$u(x,\alpha) = u_{in}(x)\, H\big(f(x,\alpha)\big) + u_{ex}(x)\, \Big(1 - H\big(f(x,\alpha)\big)\Big), \quad (4)$$

where $\alpha$ is the parameter of the level set function, $H$ is the Heaviside function.

The evolution of the level set function $f$ is performed through updating the unknown parameter $\alpha$. As a result, the level-set function $f(x)$ can be represented parametrically by using set of basis functions and unknown weights as follows:

$$f(x,\alpha) = \sum_{i=1}^{n} \alpha_i \mu_i(x) \qquad (5)$$

where $\mu = [\mu_1, \ \mu_2, \ldots\ldots, \mu_n]$ denotes our dictionary or basis vector of basis functions. Here, we used the GRBF as a basis function and it is expressed as

$$\mu_i(x) = exp(-\beta\|x - x_i\|)^2, \qquad (6)$$

where $\beta = \left(\sqrt{2}\,\sigma\right)^{-1}$, $\sigma$ is the width of Gaussian, $x_i$ is the center of GRBF, and $\|\cdot\|$ is the Euclidean norm.

## 4. Experimental results

To prove the efficacy and superiority of the proposed method in image reconstruction, we present experimental results for image reconstruction in sparse-view CT and limited-angle CT by using synthetic images and real X-ray CT projection data of a carved cheese slice [6].

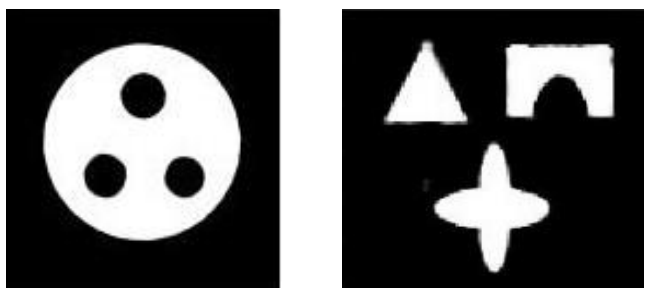

**Fig. 1**. These two synthetic images are used to evaluate the performance of the proposed approach.

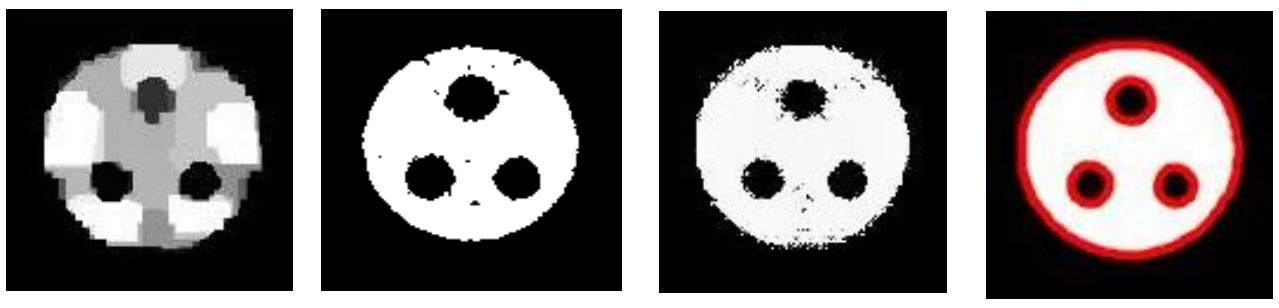

(a)TV (b)DART (c)DP (d) our method

**Fig. 2.** Reconstructed of synthetic image from only **4** projection distributed over the angular range $[0, \pi)$ compared to the other three methods.

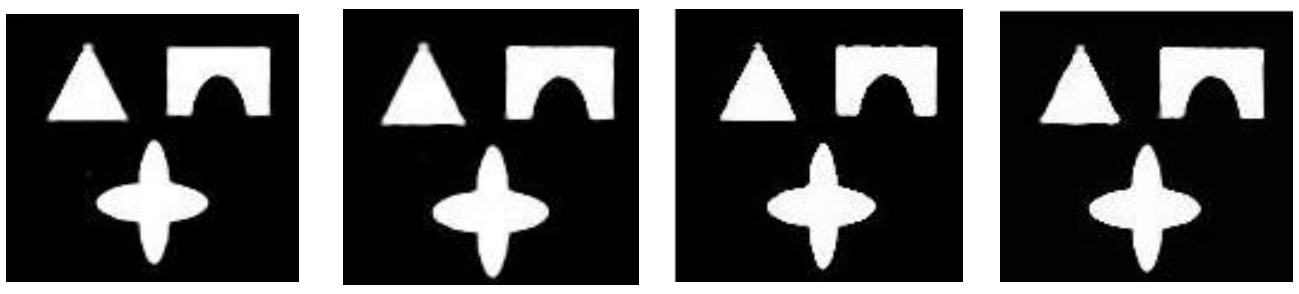

(a) $5\pi/6$ (b) $2\pi/3$ (c) $7\pi/12$ (d) $\pi/2$

**Fig. 3**. Reconstructed images in the limited-angle CT case.

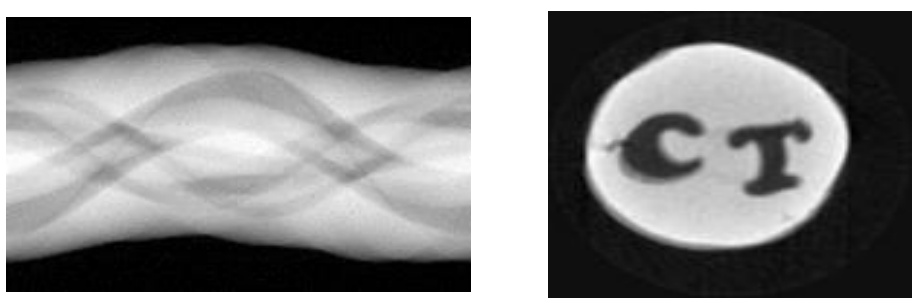

**Fig. 4.** Sinogram and a reconstructed image of the real data. with 180 angles over the angular range $[0,2\pi)$.

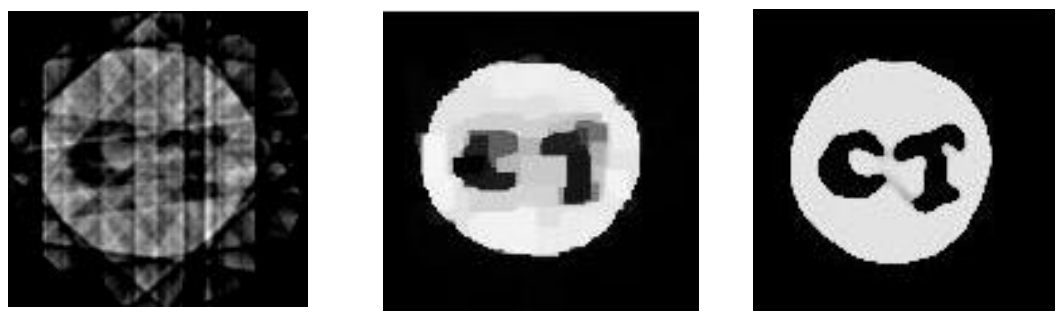

(a)FBP (b)TV (c)Our method

**Fig. 5.** Reconstructed images for real data from only **8** projection data uniformly distributed over the angular range $[0, \pi)$.

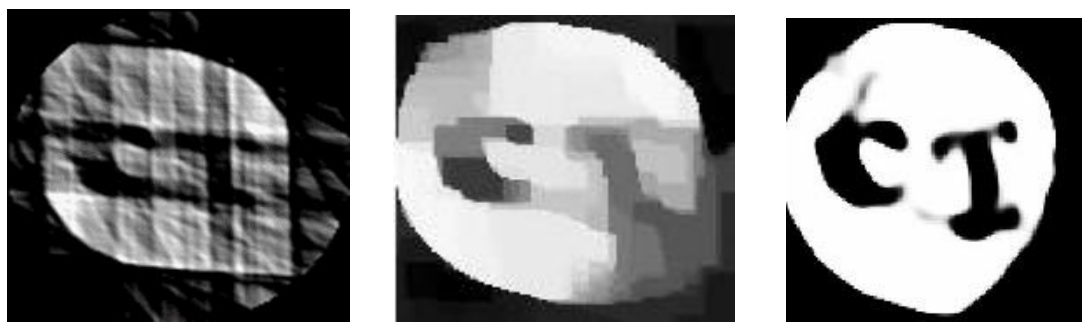

(a)FBP (b)TV (c)Our method

**Fig. 6.** Reconstructed image for the real data in the limited-angle case, where the angular range was limited to $[0, \pi/2)$.

## Conclusions

In this work, we proposed a new approach using a parametric level-set method for binary image reconstruction. We represent the object shape by using a level-set function, which in turn is represented using a linear combinations of basis vectors in the dictionary. Furthermore, the problem to be solved for image reconstruction becomes a tractable convex optimization.

## Acknowledgments

I would like to express my deep gratitude to my supervisor for his support and valuable advice in carrying out the research. This research was supported by JST-CREST Grant Number JPMJCR1765.

## Conflicts of Interest

The authors declare that they have no conflicts of interest to report regarding the present study.